\documentclass[10pt, prb, aps, amssymb, twocolumn, superscriptaddress, notitlepage, longbibliography]{revtex4-2}

\usepackage{amsmath}
\usepackage{graphicx}
\usepackage{color}
\usepackage{bbm}
\usepackage[normalem]{ulem}
\usepackage[colorlinks]{hyperref}
\usepackage{dsfont}
\usepackage{physics}
\usepackage{enumitem}
\usepackage{float}
\usepackage[english]{babel}

\newcommand{\pd}{{\phantom{\dag}}}

\begin{document}

\title{Non-Hermitian  topology of quantum spin-Hall systems to detect edge-state polarization}

\author{Raghav Chaturvedi}
\affiliation{Institute for Theoretical Physics and Astrophysics, Julius-Maximilians-Universit\"{a}t W\"{u}rzburg, Germany}
\affiliation{W\"{u}rzburg-Dresden Cluster of Excellence ctd.qmat, 01062 Dresden, Germany}
\email{raghav.chaturvedi@uni-wuerzburg.de}

\author{Ion Cosma Fulga}
\affiliation{Leibniz Institute for Solid State and Materials Research,
IFW Dresden, Helmholtzstrasse 20, 01069 Dresden, Germany}
\affiliation{W\"{u}rzburg-Dresden Cluster of Excellence ctd.qmat, 01062 Dresden, Germany}

\author{Jeroen van den Brink}
\affiliation{Leibniz Institute for Solid State and Materials Research,
IFW Dresden, Helmholtzstrasse 20, 01069 Dresden, Germany}
\affiliation{W\"{u}rzburg-Dresden Cluster of Excellence ctd.qmat, 01062 Dresden, Germany}
\affiliation{Department of Physics, TU Dresden, D-01062 Dresden, Germany}

\author{Ewelina M. Hankiewicz}
\affiliation{W\"{u}rzburg-Dresden Cluster of Excellence ctd.qmat, 01062 Dresden, Germany}
\affiliation{Institute for Theoretical Physics and Astrophysics, Julius-Maximilians-Universit\"{a}t W\"{u}rzburg, Germany}

\date{\today}

\begin{abstract}
We study the non-Hermitian topology of multi-terminal transport in a quantum
spin–Hall device described by the Bernevig–Hughes–Zhang model. We show that
breaking time-reversal symmetry alone does not imply non-reciprocal transport
or a non-Hermitian conductance matrix. Instead, non-Hermitian topology arises
only when transport becomes directionally imbalanced. We identify two distinct
mechanisms that generate such a response: spin-selective coupling at the
contacts and an out-of-plane Zeeman field that unbalances the counter-propagating
helical edge modes.  We show, for unpolarized leads, that the spin polarization–dependent response to Zeeman fields, provides a transport-based probe of the intrinsic spin polarization of the helical edge states. Moreover, we demonstrate that non-Hermitian skin effect is  more sensitive than conductance elements to detect the spin polarization of the edge states.

Our results
clarify the conditions required for non-Hermitian topology in quantum spin–Hall
transport and establish non-Hermitian skin effect as a diagnostic tool
for spin-selective coupling and edge-state polarization.
\end{abstract}

\maketitle

\section{Introduction}
\label{sec:intro}

Non-Hermitian responses in electronic transport have recently attracted attention because multi-terminal conductance matrices can exhibit features—such as asymmetric coupling between terminals, boundary-localized eigenvectors, and direction–dependent transport—that do not require non-Hermitian Hamiltonians \cite{Ochkan2024, chaturvedi2025, ozer2024, Ashida2020}. In particular, multi-terminal quantum Hall and Chern-insulator devices were shown to realize conductance matrices with the same structure as paradigmatic non-Hermitian one-dimensional chains \cite{Hatano1996, Yao2018}, displaying strongly asymmetric conductance elements and an exponential “skin” accumulation of their eigenvectors along the contact index. In those settings, the non-Hermitian structure is tied to explicitly broken time-reversal symmetry and to chiral edge states that enforce a preferred circulation direction along the sample boundary.

Whether similar behaviour can appear in time-reversal-symmetric topological systems has not been systematically explored. Quantum spin-Hall (QSH) insulators realized in HgTe/(Hg,Cd)Te quantum wells provide a natural testing ground \cite{Bernevig2006, konig2008, Konig2007}: they host helical edge states that propagate in opposite directions with opposite spin, and their transport remains reciprocal as long as the contacts do not favour one helical branch over the other \cite{Wu2006, Moore2006}. QSH devices have been extensively studied as two-dimensional topological insulators and as platforms for spin transport \cite{Roth2009, Nowack2013, Li2015, Skolasinski2018}, yet their multi-terminal conductance matrices have not been analyzed from the perspective of non-Hermitian topology. In particular, it has not been established yet whether contact-induced spin selectivity alone can generate a non-trivial non-Hermitian structure in an otherwise time-reversal-symmetric QSH system.

In this work, we show that multi-terminal QSH devices described by the Bernevig–Hughes–Zhang (BHZ) model \cite{Bernevig2006} develop a non-Hermitian topological conductance matrix when the leads couple selectively to one spin $(J_z)$ sector. This mechanism does not rely on breaking time-reversal symmetry in the scattering region and arises entirely from how the contacts inject and collect carriers from the helical edge channels. As a result, the origin of the non-Hermitian response is distinct from that in Chern-insulator devices, where chirality and magnetic fields fix the transport direction. Here, the direction and strength of the asymmetry in the conductance matrix are controlled directly by the spin polarization of the leads.

We characterize this response using complementary non-Hermitian signatures of the conductance matrix: an exponential boundary localization of its right eigenvectors along the ordered lead index (the non-Hermitian skin effect \cite{Yao2018}) and a winding number constructed from the polar decomposition of the conductance matrix \cite{Hughes2021}. These quantities evolve systematically with the lead polarization and saturate once transport is dominated by a single helical branch. In this way, non-Hermitian features of the multi-terminal conductance provide a practical tool for estimating the spin-polarization of the contacts.

We then examine how time-reversal–breaking perturbations and disorder modify
this picture. Using Zeeman fields as a controlled tuning parameter and focusing
on the case of unpolarized leads, we show that breaking time-reversal symmetry
alone does not imply non-reciprocal transport or a non-Hermitian conductance
matrix. An in-plane Zeeman field suppresses
quantized edge transport while preserving reciprocity, whereas an out-of-plane
field unbalances the counter-propagating helical edge modes and drives a
transition to a non-Hermitian topological conductance matrix. The orientation-dependent Zeeman response therefore links the emergence of
non-reciprocal transport to the intrinsic spin polarization of the helical edge
states.

Finally, we show that spin-mixing disorder can suppress the non-Hermitian skin effect and induce a crossover
from a non-trivial to a trivial conductance-matrix phase. Together, these results demonstrate that the non-Hermitian topology in QSH
transport is governed by directional imbalances between helical edge modes,
which can arise either from spin-selective contacts or edge-state polarization. As such, multi-terminal
conductance matrices provide a unified framework for analyzing spin-dependent
transport properties in QSH systems.

\section{Model and Device}
\label{sec:model}

We describe the quantum spin-Hall (QSH) system using the Bernevig--Hughes-Zhang (BHZ) model \cite{Bernevig2006, konig2008}. In the standard four-component basis 
\[
\Psi(\mathbf{k}) =
\bigl(
|E1,+\tfrac{1}{2}\rangle,\;
|H1,+\tfrac{3}{2}\rangle,\;
|E1,-\tfrac{1}{2}\rangle,\;
|H1,-\tfrac{3}{2}\rangle
\bigr)^{\mathsf T},
\]
the momentum–space Hamiltonian reads
\begin{equation}
H(\mathbf{k}) =
\begin{pmatrix}
h(\mathbf{k}) & 0 \\
0 & h^{*}(-\mathbf{k})
\end{pmatrix},
\label{eq:BHZ_continuum}
\end{equation}
with
\begin{align}
h(\mathbf{k}) &= \epsilon(\mathbf{k})\, \mathbbm{1}_{2\times2}
+ \mathbf{d}(\mathbf{k})\cdot \boldsymbol{\sigma},
\label{eq:h_block}
\\[3pt]
\epsilon(\mathbf{k}) &= C - D(k_x^{2}+k_y^{2}),
\label{eq:epsilon_def}
\\[3pt]
\mathbf{d}(\mathbf{k}) &= 
\bigl(Ak_x,\; -A k_y,\; M - B(k_x^{2}+k_y^{2})\bigr).
\label{eq:d_vector}
\end{align}

Here, the two diagonal blocks describe the two time-reversed sectors, the parameter \(A\) sets the strength of the linear coupling between the electron-like and hole-like components, \(M\) sets the system in an inverted or normal regime, and \(B,D\) regulate quadratic corrections to the dispersion.

The sign of \(M/B\) distinguishes the topological phases \cite{Bernevig2006, Wu2006}:
\(M/B<0\) corresponds to a trivial insulator, while \(M/B>0\) corresponds to a band-inverted QSH phase with a pair of counter-propagating helical edge states traversing the bulk gap \cite{konig2008, Bernevig2006}. These boundary modes are illustrated in the nanoribbon band structure of the model (finite in one direction) in the Appendix~\ref{app:ribbon}.

For numerical simulations, we discretize the model on a square lattice by replacing \(k_{x/y}\rightarrow \sin
k_{x/y}\) for linear terms and \(k_{x/y}^{2} \rightarrow 2-\cos
k_{x/y}\) for quadratic terms. The resulting tight-binding Hamiltonian on a square lattice takes the form
\begin{equation}
H = \sum_{\mathbf{r}} c_{\mathbf{r}}^{\dagger} H_0\, c_{\mathbf{r}}
+ \sum_{\mathbf{r}}
\left(
c_{\mathbf{r}+\hat{x}}^{\dagger} T_x\, c_{\mathbf{r}}
+ c_{\mathbf{r}+\hat{y}}^{\dagger} T_y\, c_{\mathbf{r}}
+ \text{H.c.}
\right),
\label{eq:BHZ_lattice}
\end{equation}
where \(c_{\mathbf{r}}\) is the four-component spinor at lattice site \(\mathbf{r}\).
The onsite term can be expressed as a diagonal $4\times4$ matrix 
\begin{equation}
H_0 = \begin{pmatrix}
h_a & 0 & 0 & 0 \\
0 & h_b & 0 & 0 \\
0 & 0 & h_a & 0 \\
0 & 0 & 0 & h_b
\end{pmatrix},
\label{eq:H0}
\end{equation}
where $h_a = M + C - 4D - 4B$ and $h_b = C - M - 4D + 4B$. The hopping terms take the form 
\begin{align}
T_x &=
\begin{pmatrix}
D+B & -A/(2i) & 0 & 0 \\
-A/(2i) & D-B & 0 & 0 \\
0 & 0 & D+B & A/(2i) \\
0 & 0 & A/(2i) & D-B
\end{pmatrix},
\label{eq:Tx}
\\[6pt]
T_y &=
\begin{pmatrix}
D+B & -A/2 & 0 & 0 \\
A/2 & D-B & 0 & 0 \\
0 & 0 & D+B & -A/2 \\
0 & 0 & A/2 & D-B
\end{pmatrix}.
\label{eq:Ty}
\end{align}

Throughout this work, we use the parameter set  $A=1$, $B=-1$, $D=-0.5$, $C=0$, $M=-1$ and set the lattice constant $a = 1$. All lengths are in units of $a$. The system lies in the inverted regime (\(M/B>0\)) and hosts a Kramers pair of helical edge states. All numerical simulations are performed using the \textsc{kwant} package \cite{Groth2014}.

We simulate a QSH square-shaped scattering region of side length $L$ and attach $N$ semi-infinite translationally invariant leads along its boundary in a clockwise sense. Here 
\begin{equation}
L = (\ell_{\mathrm{lead}} + g_{\mathrm{lead}})\,\frac{N}{4} + g_{\mathrm{lead}},
\label{eq:L_geometry}
\end{equation}
with lead width $\ell_{\mathrm{lead}} = 20$ and inter-lead spacing $g_{\mathrm{lead}} = 35$. Each lead was implemented as a semi-infinite strip of the same BHZ tight-binding model as the scattering region to ensure clean mode matching. Terminals were attached along all four edges at positions defined by the same lead-width and lead-spacing parameters as above. 

\begin{figure}[t]
\centering
\includegraphics[width=0.95\linewidth]{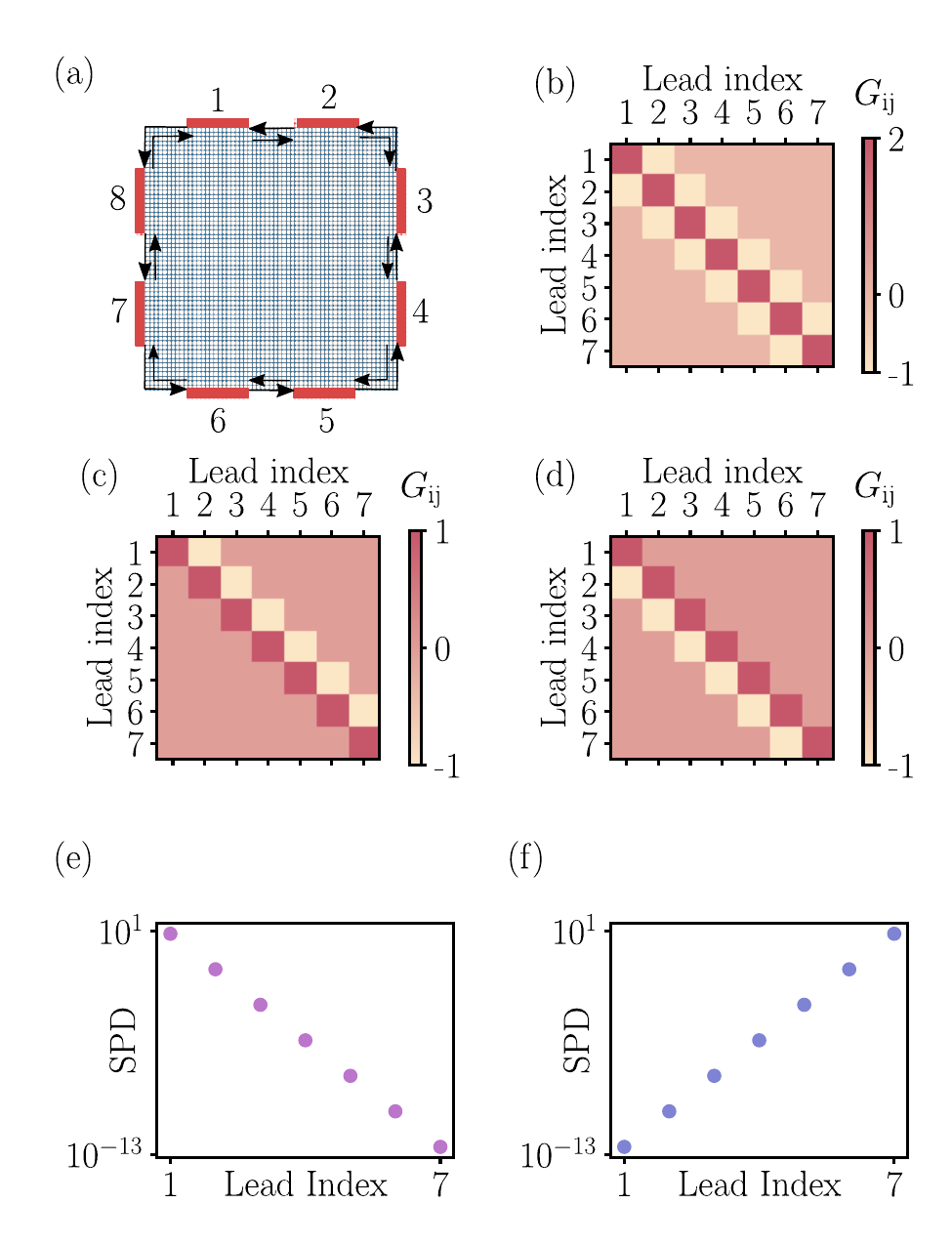}
\caption{
\textbf{Setup and emergence of non-Hermiticity with spin-polarized leads.}
(a) Square-shaped QSH device (blue) described by the BHZ model, connected to eight terminals
(red) distributed along its perimeter. The direction of helical edge states is shown using
black arrows. (b) Multi-terminal conductance matrix $G_{ij}$ for unpolarized leads;
$G$ is Hermitian, consistent with time-reversal symmetry. (c,d) Conductance matrices for
spin-polarized leads with ($\mu_{\uparrow}=+5$, $\mu_{\downarrow}= 0$) and
($\mu_{\uparrow}=0$, $\mu_{\downarrow}=5$), respectively. The asymmetry between $G_{ij}$ and
$G_{ji}$ indicates the onset of spin-selective, directional transport and renders $G$
non-Hermitian even though the underlying Hamiltonian is Hermitian. (e,f) Corresponding summed
probability density (SPD) of the right-eigenvectors of $G$, $\mathrm{SPD}=\sum_n|v_n|^2$,
plotted on a logarithmic scale along the ordered lead index. The exponential accumulation
of SPD toward opposite boundaries in the two cases shows the spin-dependent non-Hermitian
skin effect.
}
\label{fig:setup_conductance}
\end{figure}

A schematic of an eight-terminal QSH device with a well-defined lead arrangement and boundary transport is shown in Fig.~\ref{fig:setup_conductance}(a). This serves as a model for a practical device with $N$ leads connected to a system exhibiting the QSH effect, for instance, the HgTe quantum wells \cite{Konig2007}, InAs/GaSb quantum wells \cite{Knez2011}, or single layer of Bi$_2$Se$_3$ \cite{Zhang2010}.

To introduce a controllable asymmetry between the helical edge
channels, we shift the onsite energies in the leads by adding
\begin{equation}
H_{\mathrm{lead}}
= \mathrm{diag}(\mu_{\uparrow},\,\mu_{\uparrow},\,\mu_{\downarrow},\,\mu_{\downarrow})
\label{eq:lead_shift}
\end{equation}
to the Hamiltonian of each lead. Throughout this work we fix \(\mu_{\downarrow}=0\) and
vary only \(\mu_{\uparrow}\), both in units of $|M|$. Shifting the onsite energy of the $\uparrow$ block in the leads selectively changes the propagating modes belonging to $\{|E1,+\tfrac12\rangle,|H1,+\tfrac32\rangle\}$, while the
$\downarrow$ block remains unchanged. At a fixed Fermi energy, this produces an imbalance in the available channels of the two time-reversed BHZ blocks, which couple to the two counter-propagating helical edge branches. In practice, varying $\mu_{\uparrow}$ therefore realizes a spin-polarized lead in the sense that it injects and collects carriers predominantly from one of the two helical branches \cite{Brune2012}.

\section{Transport calculations}
\label{sec:transport}

Similar to previous works \cite{chaturvedi2025, ozer2024}, we perform transport simulations at zero Fermi energy \(E=0\) using the
Landauer–Büttiker formalism \cite{Buttiker1986}. In the linear regime, the currents injected and voltages measured in an \(N\)-terminal device satisfy
\begin{equation}
\vec{I} = G \vec{V}, \quad 
I_i = \sum_{j=1}^{N} G_{ij} V_j , \qquad
G_{ij} = \frac{e^2}{h}\left( \delta_{ij} N_i - T_{ij} \right),
\label{eq:LB_currents}
\end{equation}
where $\vec{I} = (I_1, \dots, I_N)^T$, $\vec{V} = (V_1, \dots, V_N)^T$ and $G$ denote the current vectors, voltage vectors, and the multi-terminal conductance matrix, respectively. Here $N_i$ is the number of modes in lead $i$ and $T_{ij}$ is the transmission probability from lead $j$ to $i$, obtained from the
scattering matrix computed using the \textsc{kwant} package \cite{Groth2014}. The constants $e$ and $h$ denote the elementary charge and Planck's constant, respectively.

By assuming the potential at the $N^{\text{th}}$ contact is set to zero, the voltages and
currents at the first $N-1$ leads can be related by the $(N-1)\times(N-1)$ submatrix of the full
conductance matrix. In the following, we denote this reduced $(N-1)\times(N-1)$ submatrix as $G$ for notational simplicity.

When the leads are unpolarized $(\mu_{\uparrow} = 0)$, time-reversal symmetry is preserved and the conductance matrix $G$ is Hermitian, as shown in Fig.~\ref{fig:setup_conductance}(b) for an 8-terminal setup.   

A finite $\mu_{\uparrow}$ creates an imbalance between the two time-reversed transport channels
at the contacts, so that one helical branch couples more strongly to the leads than the other.
As a result, $T_{ij}\neq T_{ji}$, even though the scattering region itself remains
time-reversal symmetric, and the conductance matrix becomes non-Hermitian, as seen in
Figs.~\ref{fig:setup_conductance}(c,d).
These matrices have the same structure as the Hamiltonian matrix of the Hatano--Nelson (HN)
model \cite{Hatano1996}, a simple example of a non-Hermitian tight-binding system consisting
of a fictitious chain of sites connected by asymmetric nearest-neighbor hoppings.
This analogy has previously been established in the case of conductance matrices of
time-reversal-symmetry-broken quantum Hall systems \cite{chaturvedi2025, Ochkan2024, ozer2024}, where
the off-diagonal terms of the conductance matrix are considered analogous to hoppings between
the different orbitals of the HN chain.

These non-Hermitian conductance matrices exhibit topological properties.
To see this, we label the contacts by $i$ and compute the right eigenvectors
$\{v_n\}$ of the conductance matrix $G$.
We then define the summed probability density
\begin{equation}
\mathrm{SPD}(i) = \sum_{n=1}^{N-1} |v_n(i)|^2 ,
\label{eq:SPD_def}
\end{equation}
which measures the support of the right eigenvectors of the conductance matrix on each contact
(each site of the fictitious Hatano--Nelson chain).
An exponential profile of $\mathrm{SPD}(i)$ signals the non-Hermitian skin effect, a hallmark
of non-Hermitian topology \cite{Alvarez2018, Yao2018, Hughes2021}.
Physically, this implies that in the regime where currents and voltages are proportional
($\vec{I}\propto\vec{V}$), both follow an exponentially decaying pattern between the first and
last leads.

\section{Non-Hermitian signatures as a probe of spin-selective coupling}
\label{sec:nh_signatures}

To assess the robustness and practical relevance of non-Hermitian markers of spin-selective
coupling, we evaluate all observables in this section over an ensemble of weak random
spin-independent onsite disorder realizations in the scattering region.
The disorder is introduced through a local potential
\begin{equation}
H_{\mathrm{dis}}(\mathbf{r}) = \delta(\mathbf{r}) \, \sigma_0 \otimes s_0 ,
\label{eq:anderson_disorder}
\end{equation}
where $\delta(\mathbf{r})$ is sampled independently at each lattice site from a uniform distribution $\delta(\mathbf{r})\in(-\delta,\delta]$.
This disorder does not mix spin or orbital degrees of freedom. Therefore, it preserves the
helical structure of the edge states \cite{Wu2006}.
Here, it serves solely to generate sample-to-sample fluctuations used to estimate statistical
uncertainty.

\begin{figure}[t]
\centering
\includegraphics[width=\linewidth]{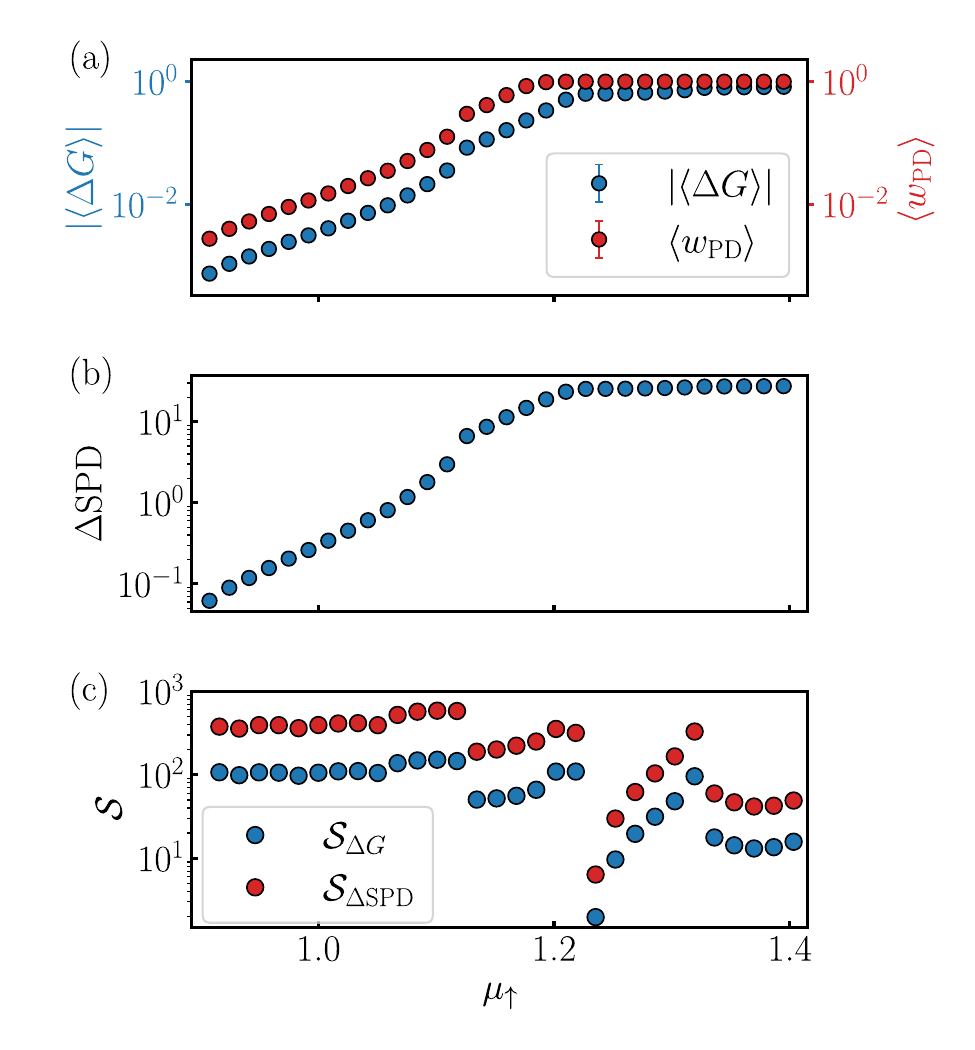}
\caption{
\textbf{Disorder-averaged transport signatures of spin-selective coupling.}
(a) Disorder-averaged adjacent-lead conductance asymmetry
$\langle \Delta G \rangle = \big|\langle G_{12}\rangle-\langle G_{21}\rangle\big|$
(left axis) and polar-decomposition invariant $\langle w_{\rm PD}\rangle$ (right axis) plotted on a logarithmic scale,
shown versus the lead-polarization parameter $\mu_{\uparrow}$ for a device with $N=32$ leads.
(b) Disorder-averaged non-Hermitian skin-effect strength $\langle \Delta \mathrm{SPD}\rangle$,
defined from the coarse-grained boundary imbalance of the summed probability density (SPD) (see Eq.~\eqref{eq:dSPDm} with $m=3$), plotted on a logarithmic scale.
(c) Noise-normalized step sensitivity $\mathcal{S}$ of $\langle \Delta G\rangle$ and
$\langle \Delta \mathrm{SPD}\rangle$, quantifying the statistical significance of their
response to incremental changes in $\mu_{\uparrow}$.
All quantities are averaged over $n_l=1000$ disorder realizations; error bars denote the
standard error of the mean; $\delta/|M| = 0.1$. Sensitivities are evaluated between adjacent polarization points $\mu_{\uparrow} $ and plotted on a logarithmic
scale.
}
\label{fig:polarization}
\end{figure}

For each disordered configuration $\alpha$, we compute the multi-terminal conductance matrix $G^{(\alpha)}$ and construct the disorder-averaged matrix: 
\begin{equation}
\langle G \rangle
=
\frac{1}{n_l}
\sum_{\alpha=1}^{n_l}
G^{(\alpha)} ,
\label{eq:Gbar}
\end{equation}
where $n_l = 1000$. Once the leads favor one of the two helical modes, the multi-terminal conductance matrix becomes
asymmetric.
A direct indicator is the charge-conductance difference between adjacent leads \cite{chaturvedi2025, Ochkan2024},
\begin{equation} \label{eq:Delta_G}
\langle \Delta G \rangle \equiv | \langle G_{12} \rangle - \langle G_{21} \rangle|,
\end{equation}
where $G_{i,j}$ denotes the total charge conductance from lead $j$ to lead $i$ (in units of
$e^2/h$).
For small values of $\mu_{\uparrow}$, this quantity remains close to zero, reflecting that both
helical branches still couple nearly equally to the contacts (see Fig.~\ref{fig:polarization} a).
Beyond a characteristic scale of $\mu_{\uparrow}$, the asymmetry grows continuously before
reaching a value of order unity (in units of $e^2/h)$, marking the crossover to a regime where one branch dominates
injection and collection processes.
Since it relies only on charge measurements, $\langle \Delta G \rangle$ provides a simple and experimentally
accessible probe of the emergence of spin-selective coupling.

A complementary non-Hermitian diagnostic is provided by the polar-decomposition invariant
$w_{\mathrm{PD}}$ \cite{Hughes2021}, which was originally introduced to characterize the topology
of fictitious one-dimensional non-Hermitian chains (the Hatano--Nelson model \cite{Hatano1996}) and has recently been extended to conductance matrices
\cite{Ochkan2024, chaturvedi2025}.
For a conductance matrix $G$, one first subtracts the average of its diagonal entries $\lambda$
to obtain $\hat{G}=G-\lambda\mathbbm{1}$, and then performs a polar decomposition
$\hat{G}=UP$, where $U$ is unitary and $P$ is positive.
The invariant is defined as
\begin{equation}
w_{\mathrm{PD}}(\hat{G}) =
\mathrm{Tr}\!\left( U^{\dagger}[U,X] \right),
\label{eq:wPD_def}
\end{equation}
where $X=\mathrm{diag}(1,2,\ldots,N-1)$ encodes the ordering of the leads and
$\mathrm{Tr}(\cdot)$ denotes the trace per unit volume evaluated over the middle of the chain.
The invariant converges to $\pm1$ in non-Hermitian topological phases and vanishes in the
trivial phase. The disorder-averaged polar-decomposition invariant is similarly defined as:
\begin{equation}
\langle w_{\mathrm{PD}}^{\pd} \rangle
=
\frac{1}{n_l}
\sum_{\alpha=1}^{n_l}
w_{\mathrm{PD}}^{\pd}(\hat{G}^{\alpha}),
\label{eq:wpd-method01}
\end{equation}
with $n_l = 1000$. The simultaneous logarithmic presentation of $\langle \Delta G\rangle$ and
$\langle w_{\rm PD}\rangle$ in Fig.~\ref{fig:polarization}(a) highlights that both quantities
respond over the same polarization window, with $\langle \Delta G\rangle$ capturing the onset
of transport non-reciprocity and $\langle w_{\rm PD}\rangle$ diagnosing the emergence of the
corresponding non-Hermitian topological phase.

While $\Delta G$ provides a scalar diagnostic, additional insight is obtained
from the summed probability density (SPD) of the right eigenvectors of the conductance matrix.
When coupling to one helical branch dominates, the eigenvector weight becomes exponentially
localized near one boundary of the fictitious Hatano--Nelson chain, reflecting the emergence of
the non-Hermitian skin effect \cite{Hatano1996, Yao2018, Alvarez2018}.
Both the direction of localization and the spatial decay length encode the sign and magnitude
of the spin-selective coupling.

A simple scalar measure of this boundary localization is the end-to-end imbalance
$\Delta \mathrm{SPD}=|\mathrm{SPD}(1)-\mathrm{SPD}(N-1)|$.
However, in finite-size and weakly disordered systems this quantity can be sensitive to local
fluctuations of individual components.
To obtain a more robust measure, we coarse-grain the boundary weight by combining several sites
and define
\begin{equation}
\Delta \mathrm{SPD}_{m} \equiv
\left|
\sum_{i=1}^{m}\mathrm{SPD}(i)
-
\sum_{i=N-m}^{N-1}\mathrm{SPD}(i)
\right|,
\label{eq:dSPDm}
\end{equation}
with $m=3$ throughout this work.
In the following, and in Fig.~\ref{fig:polarization}(b), we focus on the coarse-grained measure
$\Delta \mathrm{SPD}_{3}$ and suppress the $m$ index.

This definition averages over the first and last few leads and suppresses spurious fluctuations
without altering the underlying exponential localization.
The qualitative behaviour is insensitive to the precise value of $m$, provided $m\ll N$.
The resulting $\langle \Delta \mathrm{SPD}\rangle$ exhibits a maximum near the crossover,
reflecting saturation of boundary localization once the strongly spin-selective regime is
reached.

To quantify how sensitively a given diagnostic responds to changes in
$\mu_\uparrow$, and to enable a direct comparison between observables with
different scales and disorder-induced fluctuations, we introduce a
noise-normalized step response between adjacent polarization points labeled $v$,
\begin{equation}
\mathcal{S}_O(v) =
\frac{\left|\langle O\rangle_{v+1}-\langle O\rangle_v\right|}
{\sqrt{\mathrm{SEM}_{O,v+1}^2+\mathrm{SEM}_{O,v}^2}},
\label{eq:sensitivity}
\end{equation}
where $\langle O\rangle_v$ is any disorder-averaged observable and $\sigma_{O,v}$ is its standard deviation. The corresponding standard error of the mean is
\begin{equation}
\mathrm{SEM}_{O,v} = \frac{\sigma_{O,v}}{\sqrt{n_l}} .
\end{equation}
This quantity measures the statistical significance of the response to a
discrete change in $\mu_\uparrow$, rather than its absolute magnitude.

As shown in Fig.~\ref{fig:polarization}(c), the sensitivity
$\mathcal{S}_{\Delta \mathrm{SPD}}$ systematically exceeds
$\mathcal{S}_{\Delta G}$ over the crossover region, demonstrating that
eigenvector-based diagnostics provide a sharper and more disorder-resilient
probe of spin-selective coupling than conductance asymmetry alone.

\section{Effect of Zeeman fields on spin polarization of helical edge states}
\label{sec:Zeeman}

\begin{figure}[t!]
\centering
\includegraphics[width=0.95\linewidth]{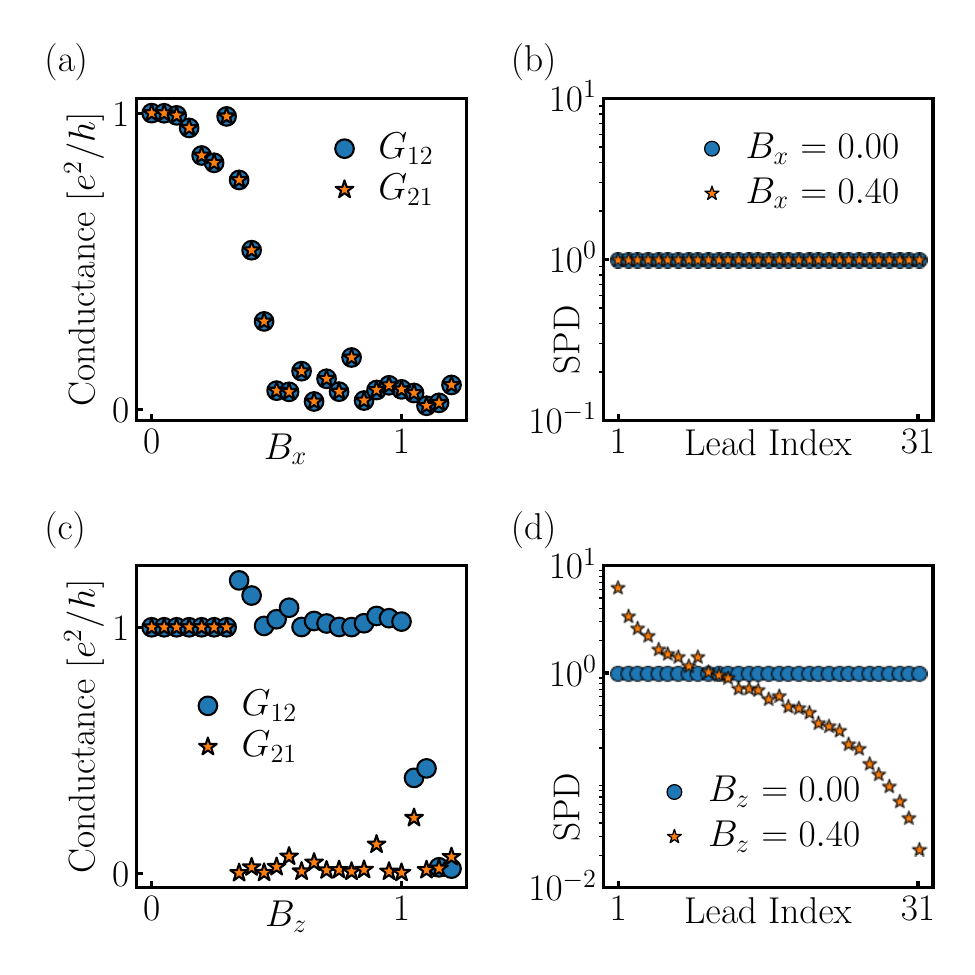}
\caption{
\textbf{Zeeman-field response for probing the spin polarization of QSH edge states.}
(a) Adjacent-lead conductances $G_{12}$ and $G_{21}$ (in units of $e^2/h$) as a
function of an in-plane Zeeman field $B_x$ for a 32-terminal device with
unpolarized leads ($\mu_{\uparrow}=\mu_{\downarrow}=0$).
Although $B_x$ breaks time-reversal symmetry and suppresses the quantized edge
conductance, reciprocity ($G_{12}=G_{21}$) is preserved.
(b) Corresponding SPD profile of the conductance matrix, which remains uniform, confirming that the
conductance matrix stays topologically trivial.
(c) Adjacent-lead conductances $G_{12}$ and $G_{21}$ (in units of $e^2/h$) for unpolarized leads under an
out-of-plane field $B_z$. Increasing $B_z$ produces pronounced non-reciprocity. (d) Corresponding SPD profile, showing the emergence of a non-Hermitian
topological phase with strong boundary localization once one helical branch is
pushed away from the Fermi level (see Appendix \ref{app:ribbon}). 
}
\label{fig:magnetic_field}
\end{figure}

Zeeman fields break time-reversal symmetry and modify the spin structure of the
helical edge states in the BHZ model. In our setup, we include the Zeeman term
\begin{equation}
H_Z = g_z B_z\, s_z \otimes \mathbbm{1} \;+\; g_x B_x\, s_x \otimes \mathbbm{1},
\label{eq:HZ_simple}
\end{equation}
where $s_{x,z}$ act in spin space, $g_x$ and $g_z$ are g-factors in x and z directions, respectively  and $B_x$, $B_z$ are given in units of $|M|$ \cite{Bernevig2006, Wu2006}. In our simulations we set $g_x = g_z = 1$.  

In the BHZ model, the helical edge states are approximately polarized along the
$s_z$ direction, with opposite $s_z$ projections associated with opposite
propagation directions \cite{Bernevig2006}. For unpolarized leads, both spin sectors couple
symmetrically to the contacts. An in-plane Zeeman field $B_x$, which does not
commute with $s_z$, mixes the two helical partners and allows backscattering
between counter-propagating modes. With increasing $B_x$, the adjacent-lead
conductance is reduced from its quantized value while remaining reciprocal,
$G_{12}=G_{21}$, and the SPD of the conductance matrix remains uniform along the
lead index (see Figs.~\ref{fig:magnetic_field}(a,b)).

An out-of-plane Zeeman field $B_z$ acts along the spin quantization axis of the
edge states and lifts the Kramers degeneracy without mixing the two helical
partners. At small $B_z$, the conductance remains reciprocal and the SPD shows no
boundary localization. With increasing $B_z$, a pronounced asymmetry between
$G_{12}$ and $G_{21}$ develops and the SPD becomes localized toward one end of the
lead index (see Figs.~\ref{fig:magnetic_field}(c,d)). This response originates
from shifting one branch of the helical edge states away from the Fermi level,
producing an imbalance between counter-propagating boundary modes
(see Appendix~\ref{app:ribbon}). 

We find that breaking time-reversal symmetry and suppressing quantized edge
transport alone are not sufficient to generate a non-Hermitian skin effect of
the conductance matrix. Instead, the appearance of the skin effect requires the
applied field to align with the intrinsic spin polarization of the helical edge
states, providing a transport-based probe of this spin polarization. Let us emphasize, that similarly to the results in Section~\ref{sec:nh_signatures}, the non-Hermitian skin effect is much more sensitive to the helical edge state polarization than elements of conductance matrix.

Finally, we note that the Zeeman-induced shift of the edge-state dispersion
(see Fig.~\ref{fig:ribbon_bands} in Appendix) provides, in principle, a route to estimate the $g_z$ factor of the helical edge modes through
$\Delta E = g_z \mu_B B_z$. This g factor of edge states is likely much smaller than the bulk, due to confinement \cite{Kernreiter2016}.

\section{Spin–mixing disorder}
\label{sec:spinmixing}

\begin{figure}[t]
\centering
\includegraphics[width=0.95\linewidth]{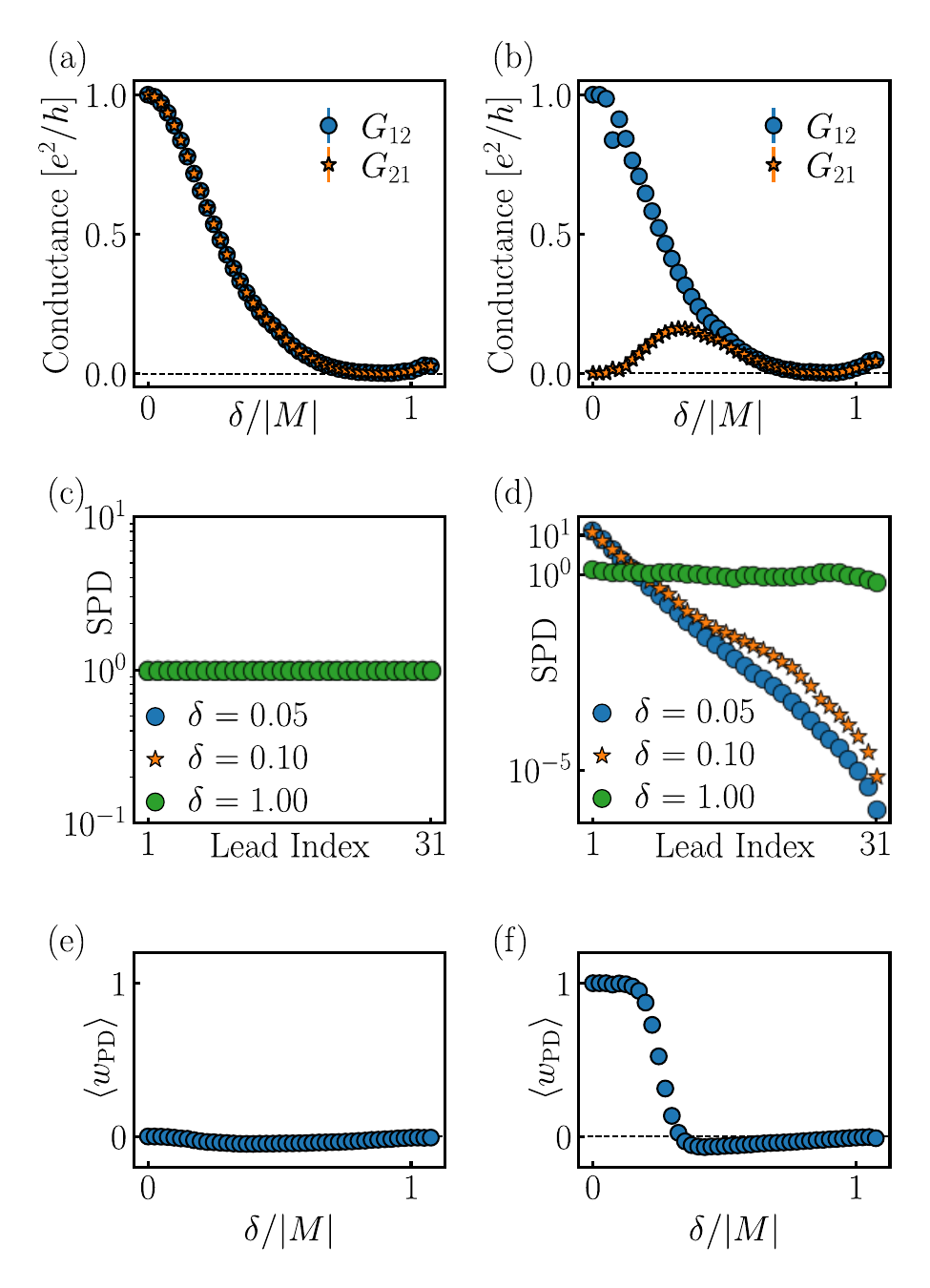}
\caption{
\textbf{Impact of spin–mixing disorder on non-Hermitian transport.}
(a,b) Disorder-averaged adjacent-lead conductances ($ G_{12} $, $ G_{21} $ in units of $e^2/h$) for a 32-terminal device
as a function of the disorder strength $\delta/|M|$ for unpolarized ($\mu_{\uparrow}=\mu_{\downarrow}=0$) and
polarized ($\mu_{\uparrow}= 4,$ $ \mu_{\downarrow}=0$) leads, respectively. The curves are obtained from the
disorder-averaged conductance matrix $\langle {G} \rangle$.  
(c,d) Corresponding summed probability density (SPD) profiles for the same
parameters, indicating the presence or absence of the non-Hermitian skin effect.
(e,f) Disorder-averaged polar-decomposition invariant $\langle w_{\mathrm{PD}} \rangle$ plotted against the disorder strength $\delta/|M|$. Panels (a,c,e)
correspond to unpolarized leads; panels (b,d,f) correspond to polarized leads.
}
\label{fig:spinflip_disorder}
\end{figure}

Spin–mixing disorder provides a second, conceptually distinct way of breaking
time-reversal symmetry without altering the orbital structure of the BHZ model.
To isolate its effect on the non-Hermitian transport response, we add to the
Hamiltonian of the scattering region, the random onsite term
\begin{equation}
H_{\mathrm{dis}}(\mathbf{r})
=
\delta(\mathbf{r})\,(s_x \otimes \mathbbm{1}),
\label{eq:H_disorder}
\end{equation}
where $\delta(\mathbf{r})\in(-\delta,\delta]$ is sampled independently at each
site and measured in units of $|M|$.  This perturbation acts only in spin
space: it locally mixes the two helical partners of the QSH edge mode and
therefore competes directly with any mechanism—such as spin-selective
contacts—that distinguishes clockwise and counterclockwise boundary propagation.

Figures~\ref{fig:spinflip_disorder}(a,c,e) show the behaviour for unpolarized
leads.  Here, both helical branches couple equally to the contacts, so the
disorder acts symmetrically on the two counter-propagating boundary modes.
Even though the conductance magnitudes decrease with increasing disorder
(Fig.~\ref{fig:spinflip_disorder}a), the adjacent-lead elements remain
reciprocal, $G_{12}\approx G_{21}$, for all $\delta$.  The SPD profiles in
Fig.~\ref{fig:spinflip_disorder}(c) remain flat across the device, and the
polar-decomposition invariant stays pinned to $\langle w_{\rm PD} \rangle =0$
(Fig.~\ref{fig:spinflip_disorder}(e)).  
Thus, for unpolarized contacts, spin–mixing disorder does not generate a
non-Hermitian skin effect or a non-trivial conductance-matrix topology,
mirroring the behaviour of an in-plane Zeeman field $B_x$ in
Fig.~\ref{fig:magnetic_field}(a,b).  Both perturbations mix the helical
partners without creating a directional imbalance, and therefore preserve a
trivial non-Hermitian phase.

For polarized contacts ($\mu_{\uparrow}\neq 0$), the clean system exhibits a
non-Hermitian response due to selective coupling to one of the helical
branches. Spin–mixing disorder counteracts this selectivity: as $\delta$
increases, the difference between adjacent-lead conductances
($G_{i,i+1}$ and $G_{i+1,i}$) steadily decreases, the SPD profile becomes
less localized, and the polar-decomposition invariant $w_{\mathrm{PD}}$
moves toward zero, as shown in Fig.~\ref{fig:spinflip_disorder} (b,d,f). This
crossover appears already for moderate disorder strengths
($\delta \approx 0.4$–$0.5\,|M|$). At this point, the disorder mixes the
two counter-propagating modes strongly enough that the contacts can no longer
distinguish them, and the conductance matrix becomes nearly reciprocal and
therefore topologically trivial. Thus, in the polarized case,
spin–mixing disorder provides a controlled route from a non-Hermitian
topological response to a trivial one at relatively modest disorder levels.

\section{Conclusions}
\label{sec:conclusion}

We showed that multi-terminal transport in quantum spin–Hall devices
described by the BHZ model may be described by a non-Hermitian conductance matrix, characterized by non-reciprocal conductance elements and a
boundary localization of conductance-matrix eigenvectors.  This establishes QSH
systems as a setting in which non-Hermitian topology can arise naturally in their transport responses, even though the Hamiltonian that governs the physical system is Hermitian.

We demonstrated that spin-selective coupling at the contacts provides one route
to such a non-Hermitian response.  Spin-polarized leads generate directional
asymmetries in adjacent-lead conductances and induce a non-Hermitian skin
effect.  We further showed that transport diagnostics based on the summed
probability density (SPD) of conductance-matrix eigenvectors and the
polar-decomposition invariant provide sensitive probes of contact polarization,
with the SPD exhibiting enhanced sensitivity to small changes in the lead
polarization.

A second route to non-Hermitian transport arises from the introduction of time-reversal–breaking magnetic fields that Zeeman couple to the electron spin.  We establish, however, that breaking time-reversal symmetry alone is not
sufficient to produce non-reciprocity: an in-plane field primarily mixes the
helical edge partners and suppresses conductance while keeping the conductance
matrix reciprocal.  By contrast, an out-of-plane field unbalances the
counter-propagating edge modes and generates a non-Hermitian topological
response even for unpolarized leads.  This establishes the orientation-dependent
Zeeman response as a probe of the intrinsic spin polarization of the helical
edge states.  Finally, we showed that spin-mixing disorder suppresses the
non-Hermitian skin effect, acting similarly to an in-plane field for
unpolarized leads and driving a crossover from a non-trivial to a trivial
conductance-matrix phase in the presence of polarized contacts.

\section{Acknowledgments}
\label{sec:ack}
We acknowledge funding by the Deutsche
Forschungsgemeinschaft (DFG, German Research Foundation) through SFB1170 ToCoTronics, Project-ID
258499086, through the Würzburg-Dresden Cluster of
Excellence on Complexity and Topology in Quantum
Matter – ctd.qmat (EXC2147, Project-ID 390858490).

\bibliography{main.bib}

\begin{appendix}

\section{Band structure}
\label{app:ribbon}

\begin{figure}[t]
    \centering
    \includegraphics[width=0.95\linewidth]{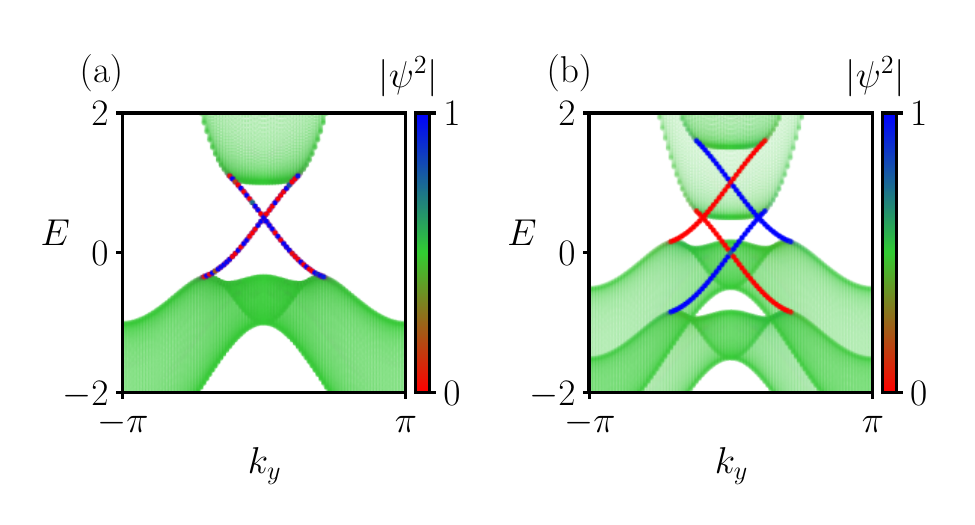}
    \caption{
        Ribbon band structure for the discretized BHZ model on a square lattice (infinite along $y$ and 100 unit cells along $x$). A color scale depicting the probability density of states $\Psi$ integrated over the first 50 unit cells is used such that states localized on the left and right ends of the ribbon are shown in blue and red, respectively. Bulk states are shown in a faded green color. (a) Zeeman field $B_z$ set to zero: helical edge states in blue and red traverse the bulk gap, with one pair on each edge. (b) $B_z=0.5$: one branch of the helical pair is pushed away from the Fermi level, leaving the opposite branch at the Fermi level ($E = 0$).}
    \label{fig:ribbon_bands}
\end{figure}

To visualize how a finite out-of-plane Zeeman field $B_z$ modifies the helical edge modes of the BHZ model \cite{Bernevig2006, Wu2006}, we compute the ribbon band structure for a system finite in the $x$-direction (100 lattice sites) and translationally invariant along $y$ so that $k_y$ remains a good quantum number. Model parameters are fixed such that the system lies in the quantum spin-Hall phase ($A=1$, $B=-1$, $D=-0.5$, $C=0$, $M=-1$). In the absence of a Zeeman field, the ribbon hosts a pair of counter-propagating helical states on each edge, which traverse the bulk gap without hybridizing (see Fig.~\ref{fig:ribbon_bands} a).

A finite $B_z$ (in units of $|M|$) breaks time-reversal symmetry and shifts the two spin blocks of the BHZ Hamiltonian in opposite directions in energy, effectively lifting the Kramers degeneracy of the edge states \cite{Bernevig2006, Brune2012}. As a consequence, one branch of the helical pair is pushed away from the Fermi energy, while the opposite branch crosses it (see Fig.~\ref{fig:ribbon_bands} b). This spectral imbalance is precisely what underlies the non-Hermitian transport response discussed in the main text: only one of the edge directions continues to contribute at zero energy, matching the behavior observed in Fig.~\ref{fig:magnetic_field}. 

\end{appendix}
\end{document}